\documentstyle[preprint,aps,epsf,floats]{revtex}

\newcommand{\LQCD}{\Lambda_{QCD}}
\newcommand{\bL}{\bar\Lambda}
\newcommand{\al}{\alpha}
\newcommand{\bt}{\beta}
\newcommand{\ga}{\gamma}
\newcommand{\eps}{\epsilon}
\renewcommand{\r}{\rho}

\newcommand{\vsl}{v\!\!\!\slash}
\newcommand{\qsl}{q\!\!\!\slash}
\newcommand{\Dp}{D_\perp}
\newcommand{\Dsl}{D\!\!\!\!\slash}
\newcommand{\Dslp}{D\!\!\!\!\slash_\perp}
\newcommand{\qv}{q\!\cdot\!v}
\newcommand{\E}{\hat{E}_0}
\newcommand{\s}{\hat{s}_0}
\newcommand{\mB}{\overline{m}_B}
\newcommand{\mD}{\overline{m}_D}
\newcommand{\G}{\Gamma}
\newcommand{\GeV}{{\rm GeV}}

\begin{document}

%\draft
 
{\tighten
\preprint{\vbox{\hbox{CALT-68-2042}
                \hbox{hep-ph/9603448} }}
 
\title{Order $1/m_b^3$ corrections to $B\to X_c\ell\bar\nu$ decay
and their implication for the measurement of $\bL$ and $\lambda_1$}
 
\author{Martin Gremm and Anton Kapustin}

\address{California Institute of Technology, Pasadena, CA 91125}

\maketitle

\begin{abstract}
We compute the order $1/m_b^3$ nonperturbative contributions to the inclusive
differential $B\rightarrow X_c\ell\bar\nu$ decay rate. They 
are parametrized by the expectation values of two local and four 
nonlocal dimension-six operators. We use our results to estimate 
part of the theoretical uncertainties in the extraction of matrix elements
$\bL$ and $\lambda_1$ from the lepton spectrum in the inclusive semileptonic
$B$ decay and find them to be very large.
We also compute the $1/m_b^3$ corrections to the moments of the hadronic
invariant mass spectrum in this decay, and combine them with the extracted
values of $\bL$ and $\lambda_1$ to put an upper bound on the branching
fraction $Br(B\rightarrow D^{**}\ell\bar\nu)$. 
\end{abstract}

}%end tighten

\newpage

\section{Introduction}

Over the last few years there has been much progress in our understanding
of the inclusive decays of hadrons containing a single heavy quark. Combining
heavy quark effective theory (HQET), with the operator product expansion
(OPE), enabled one to show that the spectator model decay rate for
$B\to X_c\ell\nu$ is the leading
term in a well-defined expansion controlled by the small parameter $\LQCD/m_Q$,
where $m_Q$ is the heavy quark mass \cite{CGG}. 
 Nonperturbative corrections to this leading approximation
 are suppressed by two powers of $m_Q$, and are parametrized by the matrix elements  
\begin{equation}
\label{l1}
\lambda_1 = \langle H_\infty(v) \,|\, \bar h_v\, (i\Dp)^2\,
  h_v\,|\, H_\infty(v)\rangle \,,
\end{equation}
and 
\begin{equation}
\label{l2}
\lambda_2 = {1\over d_H}\, \langle H_\infty(v) \,|\, \bar h_v\, {g\over2}\,
  \sigma_{\mu\nu}\, G^{\mu\nu}\, h_v\,|\, H_\infty(v)\rangle \,,
\end{equation}
where $h_v$ is the quark field in the heavy quark effective theory.
$|\,H_\infty(v)\rangle$ is the pseudoscalar ($d_P=3$) or vector ($d_V=-1$)
heavy meson state in the infinite quark mass limit
\cite{inclI,inclII,inclIII}, with normalization
$\langle H_\infty(v)\,|\,H_\infty(v^\prime)\rangle=
(2\pi)^3v^0\delta^{(3)}(p-p^\prime)$.
The scale dependent \cite{l2run} matrix element
 $\lambda_2$ can be obtained from the measured $B^*-B$ mass splitting, 
$\lambda_2(m_b)\simeq0.12\,{\rm GeV}^2$.

The determination of quantities like $\lambda_1$ and the $b$ and $c$
quark pole masses from
experiment is complicated by the presence of ultraviolet renormalons
\footnote{In the ``large $\beta_0$'' approximation $\lambda_1$ does not have a
renormalon ambiguity in continuum regularizations \cite{renorm2} but
this is likely to be an artifact of this approximation.}.
If the renormalons are present, the values of an HQET matrix element 
extracted from two different observables at a given order in $\alpha_s$ may
differ by an amount of the order of the matrix element itself\cite{renorm},
which prevents one from using
the measured value of one observable to improve the prediction
for another.
Whether this is the case can be established by expressing
the unknown HQET matrix element in terms of the first observable
and substituting this into the theoretical formula
for the second. Only if the resulting expression has a reasonably
well convergent expansion in powers of $\alpha_s$, it makes sense to
use the value of the HQET matrix element extracted from the first observable
to predict the value of the second. In practice, one knows 
only a few terms in the perturbative expansion, and it is hard to assess
how well the series converges.

Recently $\lambda_1$ and the difference between the meson masses
and the pole quark masses, $\bar\Lambda$, have been extracted from the
measured inclusive lepton spectrum in semileptonic $B$ decays \cite{Next}:
$\lambda_1=-0.19\pm0.10\GeV^2,\bL=0.39\pm0.11\GeV$.
The quoted uncertainties are the statistical errors only. There are reasons
to think that systematic experimental errors are not very large. The major
theoretical uncertainties come from order $\alpha_s^2$ perturbative corrections,
the assumption of quark-hadron duality, and the higher orders in the heavy
quark expansion. For a very similar analysis see~\cite{chern}.
An independent constraint on $\bL$ and $\lambda_1$ can be obtained from the
inclusive hadron spectrum in $B$ decays \cite{FLS}.

Here we compute the terms of order $1/m_b^3$ in the heavy quark expansion of the
differential decay rate $B\rightarrow X_c\ell\nu$ and use the results of our 
calculation to estimate part of the theoretical uncertainties in the
determination of $\bar\Lambda$ and $\lambda_1$ from inclusive $B$ decays. There are two sources of
$1/m_b^3$ corrections. First, the OPE has to be extended to include the local
dimension-six operators. Second, the lower order corrections calculated 
in Refs. \cite{inclI,inclII,inclIII} are expressed in terms of the expectation
values of dimension-five operators between the physical $B$ states, rather than 
between the states of the effective theory in the limit $m_b\rightarrow\infty$.
Therefore they depend on $m_b$ beyond leading order. 
In Sect. II we compute the contributions from the local dimension-six operators
to both the charged lepton spectrum and the
hadronic spectrum, which are experimentally accessible quantities.
The mass dependence of the states is discussed in Sect. III.
The complete $1/m_b^3$ corrections are parametrized by the expectation values
of two local and four nonlocal operators.
In Sect. IV we investigate the influence of $1/m_b^3$ corrections on the 
extraction of $\bar\Lambda$ and $\lambda_1$ from both leptonic and hadronic
spectra in $B$ decays. We also obtain an upper bound on the branching fraction
$Br(B\to D^{\ast\ast}\ell\bar\nu)$.
Our conclusions are presented in Sect. V. The Appendix
contains the derivation of the meson mass formulas to order
 $\Lambda_{QCD}^3/m_Q^2$.

\section{Local Dimension-six Operators}
The effective Hamiltonian density responsible for $b\rightarrow c\ell\bar\nu$
 decays is
\begin{equation}
\label{WeakH}
H_W=-V_{cb}\frac{4 G_F}{\sqrt 2} J^\mu J_{\ell\mu},
\end{equation}
where $J^\mu=\bar c_L\gamma^\mu b_L$ is the left-handed quark current, and
 $J_\ell^\mu=\bar\ell_L\gamma^\mu \bar\nu_L$ is the left-handed lepton current.
The differential decay rate is determined by the hadronic tensor
\begin{equation}
\label{Whad}
W^{\mu\nu}=(2\pi)^3\sum_{X_c}\delta^4\left(p_B-q-p_{X_c}\right) 
 \langle B(v)|{J^\nu}^\dagger |X_c\rangle\langle X_c|J^\mu|B(v)\rangle,
\end{equation}
which can be expanded in terms of five form factors:
\begin{equation}
\label{Formfactors}
W^{\mu\nu} = -g^{\mu\nu} W_1 + v^\mu v^\nu W_2 - 
i\eps^{\mu\nu\al\bt}v_\al q_\bt W_3
+ q^\mu q^\nu W_4 + (q^\mu v^\nu + q^\nu v^\mu) W_5. 
\end{equation}
Then the differential semileptonic decay rate is given by
\begin{eqnarray}
\label{DRate}
&&\frac{d\G}{dq^2 dE_\ell dE_\nu}={{96\,\G_0}\over m_b^5}\\ \nonumber
&& \times\left(\,W_1\,q^2 + W_2\,\left(2E_\ell E_\nu - {1\over 2}q^2\right) +
 W_3\,q^2\, (E_\ell-E_\nu)\right)\,\theta(E_\ell)\,\theta(E_\nu)\,\theta(q^2)\,
\theta(4E_\ell E_\nu - q^2).
\end{eqnarray} 
Here $\G_0$ is the spectator model total decay rate in the limit of zero charm
mass
\begin{equation}
\label{SpectRate}
\G_0 = |V_{cb}|^2 G_F^2 \frac{m_b^5}{192 \pi^3},
\end{equation}
 and we have neglected the lepton mass.

We define the current correlator $T^{\mu\nu}$ by
\begin{eqnarray}
\label{T}
T^{\mu\nu} & = & -i\int d^4x e^{-iq\cdot x}\langle B(v)|T\left[{J^\nu}^\dagger
 (x)J^\mu(0)\right]|B(v)\rangle \\ \nonumber
           & = & -g^{\mu\nu} T_1 + v^\mu v^\nu T_2 - 
i\eps^{\mu\nu\al\bt}v_\al q_\bt T_3
+ q^\mu q^\nu T_4 + (q^\mu v^\nu + q^\nu v^\mu) T_5.
\end{eqnarray}
One can easily see that $W_i=-{1\over\pi}$Im$T_i$. 
Away from the
 physical cut $T^{\mu\nu}$ can be computed using the OPE \cite{CGG}. Then
analyticity arguments show that the smeared differential decay rate is
correctly reproduced by the OPE calculation, provided the width of 
the smearing function is large enough.

\begin{figure}
\centerline{\epsfxsize=20truecm \epsfbox{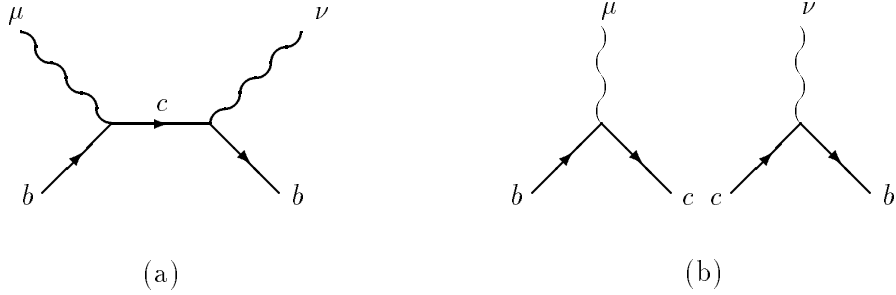} }
\caption[1]{(a) The relevant term in the operator product expansion. Wavy
lines denote the insertions of left-handed currents. (b) does not contribute
to $b\rightarrow c$ decay.}
\end{figure}

The only diagram which has a discontinuity across the physical cut is shown in
 Fig.~1a. The corresponding contribution to the time-ordered product is
\begin{eqnarray}
\label{Diagram}
&&\bar{b} \ga^\nu P_L \frac{1}{m_b\vsl - \qsl + i\Dsl - m_c}\ga^\mu P_L b
\\ \nonumber
&&= {1\over\Delta_0}\bar{b} \ga^\nu P_L \left(m_b\vsl - \qsl + i\Dsl + m_c
\right) \sum_{n=0}^{\infty} \left(\frac{D^2-2(m_b v-q)\cdot iD + {1\over 2}
g\sigma_{\al\bt} G^{\al\bt}}{\Delta_0}\right)^n \ga^\mu P_L b,
\end{eqnarray}
where $P_L={1\over 2}(1-\ga_5)$ is the left-handed
projector, $\Delta_0=(m_b v - q)^2 -m_c^2 + i0$,
 $D$ is the covariant derivative,
and we used $D_\mu D_\nu -D_\nu D_\mu = i gG_{\mu\nu}$.
The field $b(x)$ in eq.~(\ref{Diagram}) is related to the normal QCD field
by $b_{\rm QCD}(x)=e^{-im_b v\cdot x}b(x)$.
There are other contributions in the OPE of two currents, e.g., the one in
Fig.1b. However these operators do not contribute to the decay rate once
sandwiched between the $B$-meson states. For the diagram in Fig.~1b this is
ensured by $m_c$ being much larger than the available energy
in the ``brown muck,'' which is of order $\LQCD$.

Our calculation of the form factors $T_i$ follows the method of
Ref.~\cite{inclII}. We expand eq.~(\ref{Diagram}) to third
order in $D$.  The term with no derivatives is proportional to the conserved
current $\bar b\gamma_\mu b$, and thus its diagonal matrix elements can be
evaluated exactly in full QCD. All other contributions we express in terms
of the field $h_v$ in the effective theory and reexpand the
resulting expressions in powers of $1/m_b$. Therefore we need the expression
for $b(x)$ in terms of $h_v(x)$ only to order $1/m_b^2$: 
\begin{equation}
\label{HeavyF}
b(x)=\left(1+\frac{i\Dslp}{2m_b}+\frac{(v\cdot D)\Dslp}{4 m_b^2}-
\frac{\Dslp^{\,2}}{8 m_b^2}+\cdots\right) h_v(x),
\end{equation} 
where $\Dp=D-v(v\cdot D)$.
We choose to work with Foldy-Wouthuysen-type
fields, because this ensures that they satisfy the usual equal-time commutation
relations~\cite{BKP}. 

To evaluate the expectation values of the heavy
quark bilinears we need the equations of motion in the effective theory to
order $1/m_b^2$~\cite{BKP,BSUV}: 
\begin{equation}
\label{Eqmotion}
i v\cdot D h_v = \left({1\over{2 m_b}} \Dslp^{\,2} -
 {i\over{4 m_b^2}} \Dslp (v\cdot D)\Dslp+{i\over{8 m_b^2}}\left(
\Dslp^{\,2} (v\cdot D) + (v\cdot D) \Dslp^{\,2}\right)+\cdots\right) h_v.
\end{equation}
By virtue of eq.~(\ref{Eqmotion}) there are no nonperturbative corrections
to the form factors $T_i$ at order $1/m_b$~\cite{inclI}. The contributions at
order $1/m_b^2$ are expressed in terms of the matrix elements
\begin{eqnarray}
\label{redundant}
&\langle B(v)\,|\bar h_v\, (i\Dp)^2\, h_v|\,B(v)\rangle,&\\
 \nonumber
&\frac{1}{3}\langle B(v)\,|\bar h_v\, {g\over2}\,\sigma_{\mu\nu}\,
 G^{\mu\nu}\, h_v|\,B(v)\rangle &.
\end{eqnarray}
Our calculation of these contributions reproduces the results in
Ref.~\cite{inclII}. The states in the
matrix elements eqs.~(\ref{redundant}) have an implicit dependence on $m_b$.
At order $1/m_b^2$ this dependence can be neglected, in which case these
matrix elements may be replaced by $\lambda_{1,2}$ defined in eqs.~(\ref{l1})
and (\ref{l2}).
If the form factors are to be calculated to order $1/m_b^3$, this replacement
is no longer valid. An expression for the matrix elements
eq.~(\ref{redundant}) in terms of $\lambda_{1,2}$ and the expectation
values of nonlocal operators is given in Sect. III.

The $1/m_b^3$ contributions to the form factors $T_i$ from local operators can
be parametrized by
two matrix elements, $\r_1$ and $\r_2$~\cite{Mannel}.\footnote{They are related
to the matrix elements $\rho^3_D$ and $\rho^3_{LS}$ introduced in
 Ref.~\cite{BSUV} by $\r_1=\rho^3_D, \r_2={1\over 3}\rho^3_{LS}$.}
They are defined as
\begin{eqnarray}
& \langle H_\infty(v)|\bar{h}_v(iD_\al)(iD_\mu)(iD_\bt)h_v|H_\infty(v)\rangle
={1\over 3} \r_1\left(g_{\al\bt} - v_\al v_\bt \right) v_\mu, & \\
& \langle H_\infty(v)|\bar{h}_v(iD_\al)(iD_\mu)(iD_\bt)\,\gamma_\delta\gamma_5
 \,h_v|H_\infty(v)\rangle ={1\over 6}d_H\r_2 \,i\eps_{\nu\al\bt\delta}v^\nu
 v_\mu. &
\end{eqnarray}
The expectation value of any bilinear operator with three
derivatives is expressible in terms of $\rho_1$ and $\rho_2$:
\begin{eqnarray}
&\langle H_\infty(v)|\bar{h}_v\Gamma
(iD_\al)(iD_\mu)(iD_\bt)h_v|H_\infty(v)\rangle =& \\ \nonumber
&{1\over 6}\r_1\left(g_{\alpha\beta}-v_\alpha v_\beta\right)v_\mu Tr\left[
P_{+}\Gamma\right]
-{1\over 12}d_H\r_2\,i\eps_{\nu\alpha\beta\delta}
v^\nu v_\mu Tr\left[P_{+}\gamma^\delta\gamma_5 P_{+}\Gamma\right],&
\end{eqnarray}
where $P_{+}={1\over 2}(1+\vsl)$, and $\Gamma$ is any four-by-four matrix.

After a rather lengthy calculation we obtain the contributions from local
dimension-six operators to the form factors:
\begin{eqnarray}
\label{T1}
T_1^{(3)}&=&-\frac{\r_1+3\r_2}{12\Delta_0 m_b^2}
+{1\over2\Delta_0^2}\left[
\r_1-\r_2+\frac{(\r_1+3\r_2)\left(q^2-\qv^2-m_b^2+m_b\qv\right)}{3
m_b^2}\right]\\ \nonumber
&+&\frac{2(\r_1+3\r_2)}{3\Delta_0^3 m_b}(m_b-\qv)\left(
q^2-\qv^2\right)
-{4\r_1\over{3\Delta_0^4}}(m_b-\qv)^2\left(q^2-\qv^2\right),
\end{eqnarray}
\begin{eqnarray}
\label{T2}
T_2^{(3)}&=&\frac{\r_1+3\r_2}{6\Delta_0 m_b^2}+{1\over3\Delta_0^2}\left[
4\r_1+6\r_2-(\r_1+3\r_2)\,{\qv\over m_b}\right]\\ \nonumber
&+&{2\over3\Delta_0^3}\left[
(4\r_1+6\r_2)(m_b-\qv)\qv-3\r_2q^2\right]-{8m_b\r_1\over3\Delta_0^4}
(m_b-\qv)(q^2-\qv^2),
\end{eqnarray}
\begin{eqnarray}
\label{T3}
T_3^{(3)}&=&\frac{\r_1+3\r_2}{6m_b^2\Delta_0^2}\,\qv
+\frac{2(m_b-\qv)}{3\Delta_0^3}\left[(\r_1+3\r_2)\,\frac{\qv}{m_b}-3\r_2\right]
-\frac{4\r_1}{3\Delta_0^4}(m_b-\qv)(q^2-\qv^2),
\end{eqnarray}
\begin{equation}
\label{T4}
T_4^{(3)}=\frac{\r_1+3\r_2}{3m_b^2\Delta_0^2}-\frac{2\r_2}{\Delta_0^3}+
{4(\r_1+3\r_2)\over 3 m_b\Delta_0^3}\left(m_b-\qv\right),
\end{equation}
\begin{equation}
\label{T5}
T_5^{(3)}=-\frac{\r_1+3\r_2}{6m_b^2\Delta_0^2}\,\qv-
\frac{2(\r_1+3\r_2)}{3m_b\Delta_0^3}\,\qv\,(m_b-\qv)+{2\r_2m_b\over\Delta_0^3}+
\frac{4\r_1}{3\Delta_0^4}(m_b-\qv)(q^2-\qv^2).
\end{equation}

Substituting the imaginary part of these form factors into eq.~(\ref{DRate}) we
obtain the corrections to the triple differential decay rate. Interesting
quantities are the charged lepton spectrum and the hadronic spectrum.
The former is obtained by taking the imaginary part of form factors
$T_i^{(3)}, i=1,2,3$ and integrating eq.~(\ref{DRate}) over $q^2$ and $E_\nu$.
Using the rescaled lepton energy $y=2E_\ell/m_b$ we find the $1/m_b^3$
correction to the lepton spectrum
\begin{eqnarray}
\label{LeptSpect}
\frac{d\G^{(3)}}{dy}&=&{\G_0\over m_b^3}\Bigg\{\theta(1-r-y)\bigg[\,
{8\over 3}\left(3\r_1+r^2\r_1+9r^2\r_2-3r^3\r_2\right)
+{8\over 3}\r_1y(3-2r)-8\r_1y^2\\ 
\nonumber
&-&{2\over 3}\left(\r_1+3\r_2\right) y^3
-\frac{2\left(4\r_1+3r^2\r_1+9r^2\r_2\right)}{1-y}
-\frac{2r\left(8\r_1+9r\r_1+27r\r_2\right)}{3(1-y)^2}\\ \nonumber
&+&\frac{2r\left(8\r_1+17r\r_1+4r^2\r_1-9r\r_2+12r^2\r_2\right)}{3(1-y)^3}
+\frac{2r^2\left(3\r_1+4r\r_1+9\r_2+12r\r_2\right)}{(1-y)^4}\\ \nonumber
&-&\frac{8r^2\left(\r_1+3r\r_1+3r\r_2\right)}{(1-y)^5}
+\frac{40r^3\r_1}{3(1-y)^6}\bigg]-\delta(1-r-y)\frac{2(1-r)^4(1+r)^2\r_1}{3r^2}
\Bigg\},
\end{eqnarray}
where $r=(m_c/m_b)^2$. Note the contribution from the $\delta$-function at the
endpoint of the lepton spectrum. For $b\rightarrow u$ transition such singular
terms in the lepton spectrum appear already at order $1/m_b^2$, but for
$b\rightarrow c$  they do not appear until order $1/m_b^3$. This
is easily explained if one recalls that the most singular contributions to
the lepton spectrum at a given order $1/m_b^n$ can be obtained from the spectator model result
by the ``averaging'' procedure of Ref.\cite{inclII}, which involves
 differentiating $n$ times
with respect to $y$. For a massless final state quark the spectator model
spectrum has the form $f(y)\theta(1-y)$ with $f(1)\neq 0$, and thus
differentiation produces the $n-1$-st derivative of the $\delta$-function $\delta^{(n-1)}(1-y)$. For a massive quark in the final state the spectator
 model spectrum and its first derivative vanish at the end point $y=1-r$. Hence
at order $1/m_b^n$ the most singular contribution is proportional to
$\delta^{(n-3)}(1-y-r)$.

To obtain the contribution from local dimension-six operators
to the hadronic spectrum we integrate eq.~(\ref{DRate})
over $E_\nu$ and express the result in terms of rescaled hadronic variables 
$\hat{E}_0=(m_b-q\cdot v)/m_b$ and $\hat{s}_0=(m_b^2-2q\cdot v + q^2)/m_b^2$:
\begin{eqnarray}
\label{IMRate}
\frac{d \G^{(3)}}{d \s d \E } &=& \frac{8 \Gamma_0}{3m_b^3}
\Theta(\E-\sqrt{\s})\Theta(1+\s-2\E)\sqrt{\E^2-\s}\times \\ \nonumber
&& \Bigg\{
(\r_1+3 \r_2)(-3+6 \E + 2 \E^2 - 5 \s)\delta( \s-r) -2\Big[
 9(\r_1-\r_2) \\ \nonumber
&+&6(\r_1-\r_2)\s+3(\r_1+3\r_2)\s^2-(3(7\r_1-3\r_2)+11(\r_1+3\r_2)\s)\E \\
 \nonumber
&+&3((3\r_1+5\r_2)-(\r_1+3\r_2)\s)\E^2+8(\r_1+3\r_2)\E^3\Big]\delta^\prime(
 \s-r)\\ \nonumber
&-&4(\E^2-\s)\Big[3(1+\s)\r_2-(1-3\s)(\r_1+3\r_2)\E
-2(\r_1+6\r_2)\E^2\Big]\delta^{\prime\prime}( \s-r)\\ \nonumber
&+&\frac{8}{3}\E(\E^2-\s)\r_1\Big[2\s-3(1+\s)\E+4\E^2\Big]
\delta^{\prime\prime\prime}( \s-r)
\Bigg\}. \nonumber
\end{eqnarray}
The correction to the total rate is given by integrating
eq.~(\ref{LeptSpect}) or eq.~(\ref{IMRate}) over the remaining variables: 
\begin{eqnarray}
\label{TotalR}
\G^{(3)}={\G_0\over 6m_b^3}\big[\r_1(77-88r+24r^2-8r^3-5r^4+48\log r+36r^2
\log r)\\ \nonumber
+\r_2(27-72r+216r^2-216r^3+45r^4+108r^2\log r)\big].
\end{eqnarray}
The part of eq.~(\ref{TotalR}) that diverges logarithmically as $r\to 0$
agrees with the corresponding expression in Ref.~\cite{BDS}. 
There is nothing wrong with the logarithmic divergence,
 since our calculation is valid 
only for the charm mass significantly larger than $\Lambda_{QCD}$. It is the
latter condition that allowed us to discard the diagram in Fig.~1b.
For a discussion of the corrections to the total
semileptonic decay rate from dimension-six operators with a light quark in the
 final state see Ref. \cite{BDS}.

\section{Expansion of the States}
Above we have computed the $1/m_b^3$ corrections to the inclusive differential
$B$ decay rate from the local dimension-six operators in the OPE. However, there
are other sources of $1/m_b^3$ corrections.
At order $1/m_b^2$ the OPE yields the decay rate in terms of the two
matrix elements
\begin{eqnarray}
\label{l1l2}
&\langle B(v)\,|\bar h_v\, (i\Dp)^2\, h_v|\,B(v)\rangle,&\\
 \nonumber
&\frac{1}{3}\langle B(v)\,|\bar h_v\, {g\over2}\,\sigma_{\mu\nu}\,
 G^{\mu\nu}\, h_v|\,B(v)\rangle &,
\end{eqnarray}
where $|B(v)\rangle$ is the physical $B$-meson state, rather
than the state of the effective theory in the infinite mass limit
$|B_\infty(v)\rangle$. 
Thus these matrix elements are mass-dependent. At
order $1/m_b^2$ this distinction is irrelevant, but at higher orders
this mass dependence has to be taken into account explicitly.
We express the physical
states through the states in the infinite mass limit of HQET using the
 Gell-Mann and Low theorem (see e.g., Ref. \cite{FW}). 
This theorem implies that, to first order in $1/m_b$, $|B(v)\rangle$ is given by
\begin{equation}
\label{GellMannLow}
|\,B(v)\rangle=\left[1+i\int d^3x\int_{-\infty}^0dt {\cal L}_{I}(x)
-\frac{1}{V}\langle B_\infty(v)\,|i\int d^3x\int_{-\infty}^0dt {\cal L}_{I}(x)
|\,B_\infty(v)\rangle\right] |\,B_\infty(v)\rangle,
\end{equation}
where $V$ is the normalization volume and 
\begin{equation}
\label{L1}
{\cal L}_{I}={1\over {2m_b}}\bar h_v\, (i\Dp)^2\, h_v+{1\over {2m_b}} 
\bar h_v\, {g\over2}\,\sigma_{\mu\nu}\, G^{\mu\nu}\, h_v.
\end{equation}
Utilizing eq.~(\ref{GellMannLow}), one can easily expand the matrix elements 
in eq.~(\ref{l1l2}) to order $1/m_b^3$. It is convenient to introduce the
following notation:
\begin{eqnarray}\label{Tproducts}
\langle H_\infty(v)|\bar h_v\, (i\Dp)^2\, h_v i\int d^3x\int_{-\infty}^0 dt
 {\cal L}_{I}(x)|H_\infty(v)\rangle+h.c.&=&\frac{{\cal T}_1+d_H
 {\cal T}_2}{m_b}, \\ \nonumber
\langle H_\infty(v)|\bar h_v\, {g\over2}\,\sigma_{\mu\nu}\, G^{\mu\nu}\,
 h_v i\int d^3x\int_{-\infty}^0 dt {\cal L}_{I}(x)|H_\infty(v)\rangle+h.c.&=&
\frac{{\cal T}_3+d_H{\cal T}_4}{m_b}.
\end{eqnarray}
We then find
\begin{eqnarray}
\label{KbGb}
\langle B(v)\,|\bar h_v\, (i\Dp)^2\, h_v|\,B(v)\rangle
&=&\lambda_1+\frac{{\cal T}_1+3{\cal T}_2}{m_b}, \\ \nonumber
\frac{1}{3}\langle B(v)\,|\bar h_v\, {g\over2}\,\sigma_{\mu\nu}\,
G^{\mu\nu}\, h_v|\,B(v)\rangle&=&\lambda_2+\frac{{\cal T}_3+3{\cal T}_4}{3m_b}.
\end{eqnarray}
Thus these order $1/m_b^3$ corrections to the inclusive $B\rightarrow X_c
\ell\bar\nu$ decay rate
are parametrized by the matrix elements
${\cal T}_1-{\cal T}_4$ of four nonlocal operators.\footnote{These matrix elements are related to those introduced in Ref.~\cite{BSUV} as
${\cal T}_1=\rho^3_{\pi\pi}, {\cal T}_2={1\over 6} \rho^3_{\pi G},
{\cal T}_3=\rho^3_S, {\cal T}_4={1\over 3}\rho^3_A+{1\over 6}\rho^3_{\pi G}$.}

This class of $1/m_b^3$ corrections can be included in any quantity known at
order $1/m_b^2$ by using eq.~(\ref{KbGb}) to evaluate the matrix elements
of the dimension-five operators.
In particular the corrections to the form factors and the differential rates
in Ref.~\cite{inclII} can be obtained in this way.

\section{Applications}
One important application of our
results is to study the influence of $1/m_b^3$ corrections on the extraction
of the HQET matrix elements $\bL,\lambda_1$ using the methods of Refs.
\cite{Next,FLS}.

In order to compare quantities obtained from an
expansion in the inverse quark mass with experiments it is necessary to 
express the quark masses $m_c$ and $m_b$ through the physical meson masses 
$m_B$ and $m_D$ and the HQET matrix elements. 
Some details of this calculation are given in the appendix. To order $1/m_b^3$
we find the following relation
\begin{equation}
\label{mesonmass}
m_H=m_Q+\bar\Lambda-\frac{\lambda_1+d_H\lambda_2(m_Q)}{2m_Q}+\frac{\rho_1
+d_H\rho_2}{ 4m_Q^2}-\frac{{\cal T}_1+{\cal T}_3+d_H({\cal T}_2+{\cal T}_4)}{4m_Q^2},
\end{equation}
where $m_H$ is the hadron mass and $m_Q$ is the heavy quark mass.
The differential and total decay rates
are functions of the ratio of quark masses which can be expressed in terms
of the spin averaged meson masses
\begin{eqnarray}
\label{massratio}
\frac{m_c}{m_b}&=&\frac{\mD}{\mB}-\frac{\bL}{\mB}\left(1-\frac{\mD}{\mB}\right)+
\frac{\lambda_1}{2\mB^2}\left(\frac{\mB}{\mD}-\frac{\mD}{\mB}\right)
-\frac{\bL^2}{\mB^2}\left(1-\frac{\mD}{\mB}\right) \\ \nonumber
&-&\frac{\bL^3}{\mB^3}\left(1-\frac{\mD}{\mB}\right)
+\frac{\bL\lambda_1}{2\mB^3}\left(1+\frac{\mB}{\mD}-3\frac{\mD}{\mB}
+\frac{\mB^2}{\mD^2}\right)-\frac{\r_1-{\cal T}_1 -{\cal T}_3}{4\mB^3}\left(\frac{\mB^2}{\mD^2}-\frac{\mD}{\mB}\right),
\end{eqnarray}
where $\mD$ and $\mB$ are defined as $\overline{m}_{Meson}= (m_P+3m_V)/4$.

The familiar relation of the HQET matrix element $\lambda_2$ 
to the mass splitting between $B$ and $B^\ast$ mesons also needs to be extended to include the $1/m_b^3$ contributions. Using eq.~(\ref{mesonmass}) to express
the quark mass through the meson mass and $\bL$, we find
\begin{equation}
m_{H^\ast}-m_H =\Delta m_H= 2\frac{\kappa(m_Q)\lambda_2(m_b)}{m_H}
	\Big(1+\frac{\bL}{m_H}\Big)-\frac{\rho_2}{m_H^2}
	+\frac{{\cal T}_2+{\cal T}_4}{m_H^2},
\end{equation}
where $\kappa(m_Q)=(\al_s(m_Q)/\al_s(m_b))^{3/\beta_0}$ takes account of the
scale dependence of $\lambda_2$.  We can use the $B-B^\ast$ and $D-D^\ast$ mass
splitting to extract the numerical value of some of the HQET matrix elements: 
\begin{eqnarray}\label{l1extr}
\lambda_2(m_b)&=&\frac{ \Delta m_B m_B^2-\Delta m_Dm_D^2 }{2(m_B-\kappa(m_c)m_D)},
\nonumber \\
\rho_2-{\cal T}_2-{\cal T}_4 &=& \frac{\kappa(m_c)
	m_B^2 \Delta m_B ( m_D+\bL) - m_D^2 \Delta m_D ( m_B+\bL)}
	{m_B+\bL-\kappa(m_c)(m_D+\bL)}.
\end{eqnarray}

In order to extract $\lambda_1$ and $\bar\Lambda$ 
from the experimentally measured lepton energy spectrum in the
$B\to X_c\ell\bar\nu$ decay it is convenient to introduce the quantities
\cite{Next}
\begin{equation}\label{r1r2def}
R_1 = {\displaystyle \int_{1.5\,{\rm GeV}} E_\ell\,
  {{\rm d}\Gamma\over{\rm d}E_\ell}\, {\rm d}E_\ell \over \displaystyle
  \int_{1.5\,{\rm GeV}} {{\rm d}\Gamma\over{\rm d}E_\ell}\,
  {\rm d}E_\ell }\,, \qquad
R_2 = {\displaystyle \int_{1.7\,{\rm GeV}}
  {{\rm d}\Gamma\over{\rm d}E_\ell}\, {\rm d}E_\ell \over \displaystyle
  \int_{1.5\,{\rm GeV}} {{\rm d}\Gamma\over{\rm d}E_\ell}\,
  {\rm d}E_\ell }\, ,
\end{equation}
where $E_\ell$ is the lepton energy and ${\rm d}\Gamma/{\rm d}E_\ell$ is the 
complete electron energy spectrum, which we obtain by combining the results
of Ref.~\cite{inclII} with our results. In the energy spectrum at
order $1/m_b^2$, taken from Ref.~\cite{inclII}, the matrix elements
of dimension-five operators are evaluated according to eqs.~(\ref{KbGb}).
The resulting expression is combined with the contribution from local
dimension-six operators eq.~(\ref{LeptSpect}).
Expressing all quark masses through the meson masses
and using their measured values, we obtain expressions for $R_1,R_2$ in
terms of the HQET matrix elements. Combining these with perturbative corrections
and other contributions (see Ref.~\cite{Next}) we find
\begin{eqnarray}
R_1[\rm GeV] &=& 1.8059 - 0.309\,{\bar\Lambda\over\overline{m}_B}
  - 0.35\,{\bar\Lambda^2\over\overline{m}_B^2}
  - 2.32\,{\lambda_1\over\overline{m}_B^2}
  - 3.96\,{\lambda_2\over\overline{m}_B^2}
  - 0.4 {\bL^3\over\mB^3}
  - 5.7 {\bL\lambda_1\over\mB^3} \nonumber \\
&&- 6.8 {\bL\lambda_2\over\mB^3}
  - 7.7 {\rho_1\over\mB^3}
  - 1.3 {\rho_2\over\mB^3}
  - 3.2 {{\cal T}_1\over\mB^3}
  - 4.5 {{\cal T}_2\over\mB^3}
  - 3.1 {{\cal T}_3\over\mB^3}
  - 4.0 {{\cal T}_4\over\mB^3} \nonumber \\
&&- {\alpha_s\over\pi} \left(0.035+0.07\,{\bar\Lambda\over\overline{m}_B}\right)
  + \left|{V_{ub}\over V_{cb}}\right|^2
  \left( 1.33 - 10.3\,{\bar\Lambda\over\overline{m}_B} \right)
  - \left(0.0041-0.004\,{\bar\Lambda\over\overline{m}_B}\right) \nonumber\\
&&+ \left(0.0062+0.002\,{\bar\Lambda\over\overline{m}_B}\right) ,\label{R1exp}
	\\
R_2 &=& 0.6581 - 0.315\,{\bar\Lambda\over\overline{m}_B}
  - 0.68\,{\bar\Lambda^2\over\overline{m}_B^2}
  - 1.65\,{\lambda_1\over\overline{m}_B^2}
  - 4.94\,{\lambda_2\over\overline{m}_B^2}
  - 1.5\,{\bL^3\over\overline{m}_B^3}
  - 7.1\,{\bL\lambda_1\over\overline{m}_B^3} \nonumber \\
&&- 17.5\,{\bL\lambda_2\over\overline{m}_B^3}
  - 1.8\,{\rho_1\over\overline{m}_B^3}
  + 2.3\,{\rho_2\over\overline{m}_B^3}
  - 2.9 {{\cal T}_1\over\mB^3}
  - 1.5 {{\cal T}_2\over\mB^3}
  - 4.0 {{\cal T}_3\over\mB^3}
  - 4.9 {{\cal T}_4\over\mB^3} \nonumber\\
&&- {\alpha_s\over\pi} \left(0.039+0.18\,{\bar\Lambda\over\overline{m}_B}\right)
  + \left|{V_{ub}\over V_{cb}}\right|^2
  \left( 0.87 - 3.8\,{\bar\Lambda\over\overline{m}_B} \right)
  - \left(0.0073+0.005\,{\bar\Lambda\over\overline{m}_B}\right)\nonumber\\
&&+ \left(0.0021+0.003\,{\bar\Lambda\over\overline{m}_B}\right) , \label{R2exp}
\end{eqnarray}
where the first two lines contain the nonperturbative corrections to order
$1/\mB^3$.
The other terms are in order: the perturbative $\alpha_s$ corrections, the 
contribution from $B\to X_u\ell\nu$ decays, electroweak corrections, and finally
a boost, since the $B$-mesons do not decay from rest.
This is to be compared with the experimental values $R_1^{exp}=1.7831\GeV,
R_2^{exp}=0.6159$.
Neglecting the $1/\mB^3$ corrections but including statistical errors the values
$\bL=0.39\pm 0.11\GeV$ and $\lambda_1=-0.19\pm 0.10 \GeV^2$ were
found in Ref.~\cite{Next}.
 In order to take the uncertainties from the higher order
matrix elements into account, we equate the expressions for $R_{1,2}$ to 
the experimental values using $|V_{ub}/V_{cb}|=0.08,\al_s=0.22$
and eqs.~(\ref{l1extr}) to eliminate $\lambda_2$ and $\rho_2$. This yields the
extracted values of $\bL,\lambda_1$ in the form
\begin{equation}\label{inv}
\bL = f_{\bL} (R_1^{exp},R_2^{exp},\rho_1,
	{\cal T}_1,{\cal T}_2,{\cal T}_3,{\cal T}_4)
,\qquad
\lambda_1 = f_{\lambda_1}(R_1^{exp},R_2^{exp},\rho_1,
	{\cal T}_1,{\cal T}_2,{\cal T}_3,{\cal T}_4).
\end{equation}

\begin{figure}\label{regions}
\centerline{\epsfxsize=6truecm \epsfbox{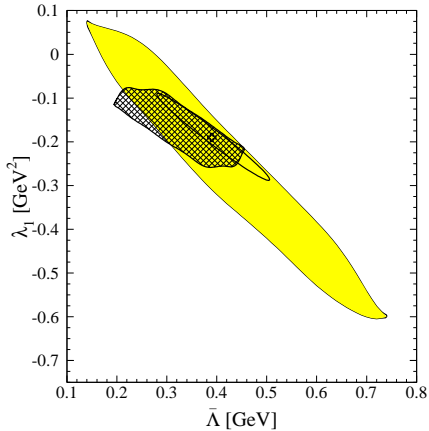} }
\caption[1]{Impact of $1/m_b^3$ corrections on the extraction of
$\bL,\lambda_1$. Shaded region: Higher order matrix elements estimated by
dimensional analysis. Cross-hatched region: $\rho_1=0.13\GeV^3,\,\rho_2=0$.
Cross and ellipse show the values of $\bL,\lambda_1$ extracted without
$1/m_b^3$ corrections but including the experimental statistical error.}
\end{figure}

Dimensional analysis suggests that the higher order matrix elements are
all of order $\LQCD^3$, which can be used to make a quantitative
estimate of the uncertainties in the extraction of $\bL,\lambda_1$. 
We vary the magnitude of $\rho_1,{\cal T}_1-{\cal T}_4$ in eqs.~(\ref{inv})
independently in the range $0-(0.5\GeV)^3$, taking $\rho_1$ to
be positive, as indicated by the vacuum saturation approximation, but making no
assumption about the sign of the other
matrix elements. Using the central values for $R_{1,2}^{exp}$ we find that
$\bL,\lambda_1$ can
lie inside the shaded region in Fig.~2. For comparison we also display
the values of $\bL,\lambda_1$ extracted in Ref.~\cite{Next} together with the
ellipse showing the size of the statistical error of the experimental data.
Clearly the theoretical
uncertainties dominate the accuracy to which $\bL,\lambda_1$ can be
extracted.

The situation can be improved only if we have some independent information
on some or all of the higher dimension matrix elements. This requires 
either more experimental input or theoretical estimates of these
matrix elements. $\rho_1$ can be estimated in the vacuum saturation
approximation\cite{SV,A9,BSUV,Mannel,CK,Next},
$\rho_1= (2\pi\alpha_s/9)m_B f_B^2$. The numerical value obtained this way
is rather uncertain. Taking $\alpha_s=0.5$ and $f_B=270$MeV for purposes of 
illustration, we find $\rho_1\simeq 0.13\GeV^3$.\footnote{
A value for $\rho_1$ can also be obtained from small velocity sum rules\cite{CK}
but this estimate suffers from large uncertainties as well.}
No similar estimates exist for the other dimension-six matrix elements.
$\rho_2$ vanishes in any non relativistic potential model, which may be taken 
as an indication that it is small relative to the other matrix elements.
No estimates that go beyond dimensional analysis are available for the time
ordered products.

The cross hatched region in Fig.~2 shows the range
of $\bL,\lambda_1$ one obtains from setting $\rho_1=0.13 \GeV^3$ and $\rho_2=0$
and varying the magnitude of the other matrix elements in the range
$0-(0.5\GeV)^3$. The previously extracted values of $\bL,\lambda_1$ are not
excluded by this choice of $\rho_{1,2}$.

This method of extracting $\bL,\lambda_1$ is especially sensitive to higher
order corrections since the constraints obtained form $R_1$ and $R_2$ give
almost parallel bands in the $\bL-\lambda_1$ plane. Thus small uncertainties
in the theoretical expressions for $R_{1,2}$
result in large uncertainties in the extracted
values of $\bL,\lambda_1$. The same applies to the very similar analysis in
Ref.~\cite{chern}. The rare decay $B\to X_s\gamma$ provides a way to
extract a vertical band in the $\bL-\lambda_1$ plane, but at present the
experimental data does not allow a quantitative analysis\cite{lig}.
Furthermore, as discussed in the introduction, it is not clear when HQET
matrix elements extracted from different observables can be compared
meaningfully\cite{renorm,FLS}.

The second method for extracting information on $\bL,\lambda_1$ \cite{FLS}
was used to exclude some regions in the $\bL-\lambda_1$
plane. The first and second 
moments of the invariant mass spectrum of the hadrons in the
final state of the inclusive decay $B\to X_c\ell\nu$ turn out to give
independent constraints on $\bL,\lambda_1$. Their definition involves the 
total decay rate at order $1/\mB^3$. It can be
obtained by combining the total rate
at order $1/m_b^2$ from Ref.~\cite{inclII} with the contributions from local
dimension-six
operators eq.~(\ref{TotalR}) and using eqs.~(\ref{KbGb}).
Finally eqs.~(\ref{mesonmass}) and (\ref{massratio}) are used 
to eliminate the quark masses.
Using the measured values for the meson masses and neglecting perturbative
corrections we find to third order in $1/\mB$:
\begin{eqnarray}\label{numrate}
\Gamma&=&\frac{|V_{cb}|^2G_F^2\mB^5}{192\pi^3} \Bigg[
0.3689-0.6080\frac{\bar\Lambda}{\mB}-0.349
\frac{\bar\Lambda^2}{\mB^2}-1.175\frac{\lambda_1}{\mB^2}-2.757
\frac{\lambda_2}{\mB^2}-0.11\frac{\bar\Lambda^3}{\mB^3} 
\\ \nonumber 
&-&1.21\frac{\bar\Lambda \lambda_1}{\mB^3}
+2.95\frac{\bar\Lambda\lambda_2}{\mB^3}-2.27\frac{\r_1}{\mB^3}+
2.76\frac{\r_2}{\mB^3}-2.73\frac{{\cal T}_1}{\mB^3}+0.55\frac{{\cal
 T}_2}{\mB^3}-3.84\frac{{\cal T}_3}{\mB^3}-2.76\frac{{\cal T}_4}{\mB^3} 
\Bigg] . 
\end{eqnarray}
Since none of the coefficients of the higher order matrix elements turn out
to be abnormally large, dimensional analysis indicates that the $1/\mB^3$
corrections to the total rate should not exceed 2\%.

The hadronic moments are defined as
\begin{equation}
\langle (s_H-\mD^2)^n\rangle=\frac{1}{\G}\int {\rm d}s_H{\rm d}E_H (s_H-\mD^2)^n
	\frac{ {\rm d}\G}{{\rm d}s_H{\rm d}E_H},
\end{equation}
where $s_H=m_B^2-2m_B v\cdot q + q^2$ and $E_H = m_B - v\cdot q$ are the
hadronic analogs of $\s,\E$ defined in Sect. II.
Using the relation between quark and hadron masses one can 
relate $s_H,E_H$ to $\s,\E$ and thus compute the moments using the
expressions given in Ref.~\cite{FLS} together with eq.~(\ref{IMRate}) and the
usual substitution eqs.~(\ref{KbGb}).
We find to order $1/\mB^3$:
\begin{eqnarray}\label{hmoms}
\langle s_H-\mD^2\rangle &=& \mB^2 \bigg[ 0.051 \frac{\alpha_s}{\pi} 
	+ 0.23 \frac{\bar \Lambda}{\mB}\left(1+0.43\frac{\alpha_s}{\pi}\right)
	+ 0.26\frac{1}{\mB^3}(\bL^2+3.9\lambda_1-1.2\lambda_2) \nonumber \\
&&+ 0.33\frac{1}{\mB^3}( \bL^3 + 6.6\bL\lambda_1 -1.7 \bL \lambda_2
		+ 7.0 \rho_1 + 3.5\rho_2 \nonumber \\
&&+ 5.0 {\cal T}_1 + 2.5 {\cal T}_2 + 4.6{\cal T}_3+1.3{\cal T}_4)
	\bigg] \\
\langle (s_H-\mD^2)^2\rangle &=& \mB^4 \bigg[ 0.0053 \frac{\alpha_s}{\pi} 
	+ 0.038 \frac{\bar \Lambda}{\mB}\frac{\alpha_s}{\pi}
	+ 0.065\frac{1}{\mB^3}(\bL^2-2.1\lambda_1) \nonumber \\
&&+ 0.14\frac{1}{\mB^3}( \bL^3 + 2.2 \bL\lambda_1 + 2.2 \bL\lambda_2
		- 6.0 \rho_1 + 1.7\rho_2
		- 1.0 {\cal T}_1 - 2.9 {\cal T}_2 )
	\bigg]
\end{eqnarray}
where perturbative $\alpha_s$ corrections have been included.
Rather that repeating the analysis presented in Ref.~\cite{FLS}, we use these
expressions to predict the values of the hadronic moments using the 
HQET matrix elements extracted from the lepton energy spectrum. The main 
reason for doing this is that the experimental measurement of the necessary 
branching fractions is not very precise.
In particular ALEPH and CLEO quote only an upper bound for 
$Br(B\to D^{\ast}_2\ell\bar\nu)$ \cite{aleph,cleo}.
We extract an upper bound on this branching fraction from
the theoretical prediction of the hadronic moments.
A lower bound for the first hadronic moment is given by \cite{FLS}
\begin{eqnarray}\label{bound1}
\langle s_H-\mD^2\rangle &\ge& a\Bigg[(2.450\GeV)^2-(1.975\GeV)^2\Bigg] 
	+ b\Bigg[(2.010\GeV)^2-(1.975\GeV)^2\Bigg]  \nonumber \\
&&	+ c \Bigg[ (1.869\GeV)^2-(1.975\GeV)^2\Bigg]
\end{eqnarray}
where $a,b$, and $c$ are the semileptonic 
branching fractions to $D^{\ast\ast},D^\ast$,
and $D$ relative to the total semileptonic branching fraction.
Using the measured ratio 0.41:0.59 for the decays to $D$ and
$D^\ast$, we can write $b,c$ as functions of the branching fraction $a$ for
$D^{\ast\ast}$:
\begin{equation}
b = 0.59(1-a) , \qquad c = 0.41(1-a) 
\end{equation}
We take $a+b+c=1$, which is appropriate because we need only a lower
bound on the hadronic moment. It is also implicitly assumed that
the nonresonant semileptonic branching fraction below the $D^{\ast\ast}$ mass
is negligible.
Similarly, for the second hadronic moment we take
\begin{equation}\label{bound2}
\langle (s_H-\mD^2)^2\rangle \ge a\Bigg[ (2.450\GeV)^2-(1.975\GeV)^2\Bigg]^2,
\end{equation}
where small contributions from the ground state mesons $D,D^\ast$
 have been neglected.
We obtain theoretical predictions for the hadronic moments by substituting 
values of $\rho_1,{\cal T}_1-{\cal T}_4$ and the corresponding values
of $\bL,\lambda_1$ extracted from the lepton spectrum into eqs.~(\ref{hmoms}).
As before, we allow the magnitudes of $\rho_1$ and ${\cal T}_1-{\cal T}_4$
to vary in the range $0-(0.5\GeV)^3$ with $\rho_1$ being positive.
Imposing the constraint that the largest values of the hadronic moments
obtained from this procedure be larger than the lower bounds 
eqs.~(\ref{bound1}),(\ref{bound2})
we find the upper bound on the $D^{\ast\ast}$ branching fraction
\begin{equation}
a \le 0.23\, .
\end{equation}
This value is compatible with the experimentally measured values
from ALEPH \cite{aleph}
($Br(B\to D_1\ell\nu)=0.069\pm0.015, Br(B\to D^\ast_2\ell\nu)<0.11$)
and from CLEO\cite{cleo}($Br(B\to D_1\ell\nu)=0.046\pm0.013,
Br(B\to D^\ast_2\ell\nu)<0.11$).
It is also marginally consistent with the OPAL result $a=0.34\pm 0.07$\cite{opal}.
Unless the matrix elements of dimension-six operators
are even bigger than we have assumed,
this implies that the branching fraction $a=0.27$ used in \cite{FLS}
is inconsistent with the values of $\bL,\lambda_1$ extracted from the
lepton spectrum.

\section{Conclusions}

We have calculated the $1/m_b^3$ contributions to various observables in
the semileptonic decay $B\to X_c\ell\nu$. They are parametrized by the
expectation values of two
local and four nonlocal dimension-six operators.  While the total rate is rather
insensitive to the higher order corrections (1-2\%), the values of
$\bL,\lambda_1$ extracted from the lepton spectrum
can be affected substantially.
The theoretical uncertainties in the values of $\bL,\lambda_1$ 
are far larger than the statistical errors of the experimental measurements
if the values of the higher order matrix elements are estimated using
dimensional analysis. While one linear combination of $\bL$ and $\lambda_1$ is
still reasonably well constrained, it is not possible to extract individual
values for $\bL$ and $\lambda_1$ from the lepton spectrum only.
The situation can be improved only if additional information on the size
of the dimension-six matrix elements is used. Unfortunately no
theoretical estimates are available for any of these matrix elements except
$\rho_1$. The latter can be estimated in the vacuum saturation approximation,
albeit with large uncertainties.
Alternatively one can use additional experimental input, e.g., from
$B\to X_s\gamma$ decays, to further constrain $\bL$ and $\lambda_1$\cite{lig}.

The values of $\bL,\lambda_1$ extracted from the lepton spectrum can be
 used to make theoretical
predictions for the moments of the hadronic invariant mass spectrum.
This amounts to expressing one observable in terms of other observables,
a procedure that makes sense only if the perturbative series for this
expression is reasonably well behaved. In order to determine whether this
is the case it is necessary to know at least the next-to-leading order $\al_s$
corrections to all observables involved. Since they have not been computed
for the lepton spectrum, there is at present no way we
can check if predictions for the hadronic moments in terms of the HQET
matrix elements extracted from the lepton spectrum satisfy this criterion.

Setting these considerations aside, we can predict the values of the 
hadronic moments in terms of $\bL,\lambda_1$ extracted from the lepton
spectrum.  The lower bounds for these moments 
depend on the branching fraction to $D^{\ast\ast}$, which 
is not well known experimentally.
By demanding that not the whole range of predicted values of the 
hadronic moments be excluded by the lower bounds, we find
an upper bound of 23\% on the branching fraction to
$D^{\ast\ast}$, if the higher order matrix elements are estimated by
dimensional analysis. This value is consistent with the ALEPH and CLEO
measurements.

\acknowledgments
We are grateful to Zoltan Ligeti and Mark Wise for helpful discussions.
This work was supported in part by the U.S.\ Dept.\ of Energy under Grant no.\
DE-FG03-92-ER~40701. A.\ K.\ was also supported by the Schlumberger
 Foundation.

\appendix
\section{The Mass Formula}
For comparison with experiments it is necessary to express the pole quark
masses $m_c$ and $m_b$ in terms of HQET matrix elements and physical
observables, e.g., the spin averaged meson masses $\mB$ and
$\mD$, where $\overline{m}_{Meson}= (m_P+3m_V)/4$. For this purpose
one needs to know how quark masses are related to hadron masses at order
$1/m_b^3$.
Our starting point is the identity
\begin{equation}
\label{Id1}
m_H=\frac{V}{2}\frac{\langle H_\infty(v)\,|{\cal H}|\,H(v)\rangle}{\langle
 H_\infty(v)\,|\,H(v)\rangle} + h.c.\, ,
\end{equation}
where $V$ is the normalization volume and ${\cal H}$ is the full Hamiltonian
density including light degrees of
freedom\footnote{If one starts from a similar identity with eigenstates of
 ${\cal H}$ on both sides of the matrix element, one obtains the same result
 after a somewhat more cumbersome calculation.}. This equation holds in the
rest frame of the hadron.
Then we split $\cal H$ into the leading term and the terms suppressed by powers
 of $1/m_b$, ${\cal H}={\cal H}_0+{\cal H}_1$, and use the fact that
$|\,H_\infty(v)\rangle$ is an eigenstate of $\int d^3x{\cal H}_0$ with eigenvalue
$m_b+\bar\Lambda$. The use of the Foldy-Wouthuysen-transformed fields
eq.~(\ref{HeavyF}) ensures that there is no implicit dependence on $m_b$
in $h_v$. Also, there are no time-derivatives in the HQET Lagrangian
beyond leading order, as can be seen e.g. from eq.~(82) of Ref.~\cite{BKP}.
Therefore we have ${\cal H}_1 = -{\cal L}_1$.
 Using the Gell-Mann and Low theorem,
 the general expression for the hadron mass reads
\begin{equation}
\label{Id2}
m_H=m_b+\bar\Lambda-\frac{V}{2}\Bigg[ \frac{\langle
H_\infty(v)\,| {\cal L}_1\,
T\exp\left(i\int d^3x\int^{0}_{-\infty}dt{\cal L}_1(x)\right) 
|\,H_\infty(v)\rangle}{\langle
H_\infty(v)\,|T\exp\left(i\int d^3x\int^{0}_{-\infty}dt{\cal L}_1(x)\right)
 |\,H_\infty(v)\rangle} + h.c. \Bigg].
\end{equation}
Expanding eq.~(\ref{Id2}) to order $1/m_b^3$ we obtain the mass formula:
\begin{eqnarray}
\label{Id3}
m_H&=&m_b+\bar\Lambda-\langle H_\infty(v)\,|
{\cal L}_I+{\cal L}_{II}|\,H_\infty(v)\rangle \\ \nonumber
&&-\left[\,{1\over 2}\langle H_\infty(v)\,|{\cal L}_Ii
\int d^3x\int_{-\infty}^0dt{\cal L}_I(x)|\,H_\infty(v)\rangle+h.c.\,
\right],
\end{eqnarray}
where~\cite{BKP}
\begin{equation}
{\cal L}_{II}= -{1\over{4m_b^2}}\bar h_v\, i\Dslp (iv\cdot D) i\Dslp\,h_v
+{1\over{8m_b^2}}\bar h_v\, (i\Dslp)^2 (iv\cdot D)\, h_v
+{1\over{8m_b^2}}\bar h_v\,  (iv\cdot D) (i\Dslp)^2\, h_v,
\end{equation}
and ${\cal L}_I$ is given in eq.~(\ref{L1}).
Eq.~(\ref{Id3}) contains expectation values of both local and nonlocal
operators. The local part can be evaluated in terms of the matrix elements
$\lambda_1,\lambda_2,\rho_1$ and $\rho_2$, while the nonlocal matrix elements
 can be expressed through ${\cal T}_1-{\cal T}_4$ defined in
 eqs.~(\ref{Tproducts}).  In terms of these matrix elements the meson mass
is given by
\begin{equation}
m_H=m_b+\bar\Lambda-\frac{\lambda_1+d_H\lambda_2}{2m_b}+\frac{\rho_1+d_H\rho_2}{
4m_b^2}-\frac{{\cal T}_1+{\cal T}_3+d_H({\cal T}_2+{\cal T}_4)}{4m_b^2},
\end{equation}
in agreement with Refs.~\cite{BKP,BSUV}.

%\newpage

\end{document}